\documentclass[12pt]{iopart}
\usepackage{graphicx}

\newcommand{\z}{Zr$_2$Ru$_3$Si$_4$}
\newcommand{\hct}{\ensuremath{H_{\rm c2}(T)}}

\newcommand{\kb}{\ensuremath{k_{\it B}}}
\newcommand{\tc}{\ensuremath{T_{\it c}}}

\begin{document}

\title{Superconducting properties of the ternary transition-metal silicide \z}

\author{Soshi~Ibuka, Motoharu~Imai, Takashi~Naka and Mitsuaki~Nishio}
\ead{IBUKA.Soshi@nims.go.jp}
\address{National Institute for Materials Science, Tsukuba, Ibaraki 305-0047, Japan}

\begin{abstract}
Superconducting properties of the polycrystalline {\z} were investigated by the electrical resistivity, magnetization and specific heat. 
By these measurements, bulk superconductivity with transition temperature $\tc = 5.5$~K was confirmed. 
Moreover, {\z} was found to be a type-II and intermediate-coupling superconductor. 
Interestingly, the electronic specific heat shows a deviation from a one-gap $s$-wave model and {\hct} shows unusual positive curvature in the vicinity of {\tc}. The first principle's calculation shows the existence of plural anisotropic Fermi surfaces. These results suggest that {\z} is not an isotropic single-gap superconductor, but possibly a multi-gap or an anisotropic gap superconductor.
\end{abstract} 

\maketitle

\section{Introduction}
Superconductivity in the ternary silicides $R_xT_y$Si$_z$ containing a rare-earth element $R$ and a transition element $T$ was explored with an expectation of finding new examples of interactions between superconductivity and magnetism at the end of 1970s and in 1980s~\cite{Bra84}. In this period, superconductors with various attractive features have been discovered, such as the first heavy fermion superconductor (CeCu$_2$Si$_2$~\cite{Ste79}), multi-gap superconductors (Lu$_2$Fe$_3$Si$_5$\cite{Bra80, Tam08} and Sc$_5$Ir$_4$Si$_{10}$~\cite{Bra80b, Tam08}) and a noncentrosymmetric superconductor (LaPtSi~\cite{Kle82, Bra84, Eve84, Kne13}). Reflecting these successful results, the efforts for searching new superconductors in $R_xT_y$Si$_z$ still have been continued until now.

Ternary Zr-Ru-Si system has not yet been fully investigated and has been reported only in two compounds: ZrRuSi~\cite{John72} and {\z}~\cite{Cha85}. Their physical properties are little known, either. An ambient phase of ZrRuSi synthesized by arc-melting has been reported to show no superconductivity down to 1.2~K~\cite{Bar80, Zho86}, while that synthesized at a high pressure shows it with {\tc} of 7--12~K~\cite{Shi95}, although it is isostructural to the ambient phase. Additionally, another high-pressure phase of ZrRuSi has been reported to be a superconductor with {\tc} of 3--5~K~\cite{Shi95}. As for {\z}, superconductivity with {\tc} of 5.6~K was originally reported in a half page of abstract for a presentation by H. F. Braun {\etal} in 1986~\cite{Bra86, lb}, but no detail has been reported so far.
{\z} has the Hf$_2$Ru$_3$Si$_4$-type structure (space group $C2/c$, monoclinic) with $a = 19.0$, $b = 5.34$, $c = 13.3$~{\AA} and $\beta = 127.73^{\circ}$~\cite{Cha85}. The structure is characterised by infinite columns of face-shared Ru-centred Si-octahedra and infinite columns of face-shared Si-centred square antiprisms~\cite{Cha85b}.
Three compounds, Hf$_2$Ru$_3$Si$_4$~\cite{Cha85}, Yb$_2$Ru$_3$Ge$_4$~\cite{Fal07} and {\z}~\cite{Cha85}, are known to crystallise in this structure; only {\z} was reported to be a superconductor above 1.8~K among the three compounds. 

Thus, in this study, we investigated the superconducting properties of the polycrystalline {\z} by electrical resistivity, magnetization and specific heat measurements.
From these measurements, we established bulk superconductivity with $\tc = 5.5$~K. The field dependence of the magnetization characterises it as type-II superconductivity. Additionally, specific heat and {\hct} results suggest that {\z} is not a simple $s$-wave superconductor, but likely a multi-gap or an anisotropic gap superconductor, attributed to the low-symmetric structure. 

\section{Material and methods} 
Polycrystalline {\z} was prepared by arc melting and subsequent annealing. The starting materials of 99.9\% purity for Zr, 99.9\% for Ru and 99.99999999\% for Si 
were weighted with the molar ratio of Zr:Ru:Si = 20:35:45 and melted on a water-cooled copper hearth under high purity argon gas atmosphere. The proportion of Ru and Si to Zr was increased from the stoichiometric composition 2:3:4 to decrease ZrRuSi impurity. The resulting as-cast alloy was put in an alumina crucible, sealed in a quartz tube under a pure Argon atmosphere, and heat-treated at 1273~K for 100~h in an electric furnace. The sample was cooled down in the furnace by turning off the heater. 
The annealed sample was characterised by X-ray powder diffraction (RINT TTR-III, Rigaku) with Cu K$\alpha$ radiation and electron probe microanalyzer (JXA-8500F, JEOL). Atomic compositions were determined by wave dispersive X-ray spectroscopy. 
The sample was cut into rectangles for each measurement.
A four-probe electrical resistivity measurement with an ac current density of 1 mA/mm$^2$ at 1 Hz in magnetic field transverse to the current were performed down to $T = 1.8$~K 
with a commercial instrument, physical properties measurement system (PPMS, Quantum Design). Heat capacity was measured by a thermal relaxation method on a PPMS.
A dc magnetization measurement was carried out with a commercial superconducting quantum interference device magnetometer (MPMS, Quantum Design). Pressure dependence of {\tc} up to 1.7 GPa was investigated by using 
a piston-cylinder-type clamp cell with a NiCrAl liner and a CuBe outer cylinder~\cite{Erem96}.
A four-probe electrical resistivity measurement was employed by using PPMS.
The sample was mounted on a specially designed plug and inserted into a Teflon cell with Daphne 7373 (Idemitsu Kosan Co.) pressure-transmitting media.
Generated pressure in the cell was calibrated against the load using {\tc} of Pb~\cite{Bire88}.
The temperature at the sample position was measured using an extra calibrated thin film resistance sensor (Cernox, Lake Shore Cryotronics). 
Two sets of the measurements were performed from the low to high pressures with using the same sample. 
Additionally, band structure, electronic density of states (DOS) and Fermi surfaces of {\z} were calculated by the DFT-based plane-wave basis sets and ultrasoft pseudopotentials method~\cite{Van90} implemented in Quantum-ESPRESSO package~\cite{qe}. 
The plane-wave energy cutoff was set to 25 Ry.
The structural parameters of {\z} were fixed at the experimentally observed values in \cite{Cha85}.
Zr ($4s, 4p, 4d, 5s, 5d$), Ru ($4d, 5s, 5p$) and Si ($3s, 3p$) were treated as valence electrons. For the exchange-correlation functional, the Perdew-Burke-Ernzerhof semi-local density generalised-gradient approximation~\cite{Per96} was used. 
The Brillouin zone was sampled with Monkhorst-Pack $20 \times 10 \times 20$ $\vec{k}$-point grids~\cite{Mon76}. The total energy convergence was confirmed to be within 3~meV/atom.

\section{Results}
\begin{figure}
	\begin{center}
	\includegraphics[width=0.57\hsize]{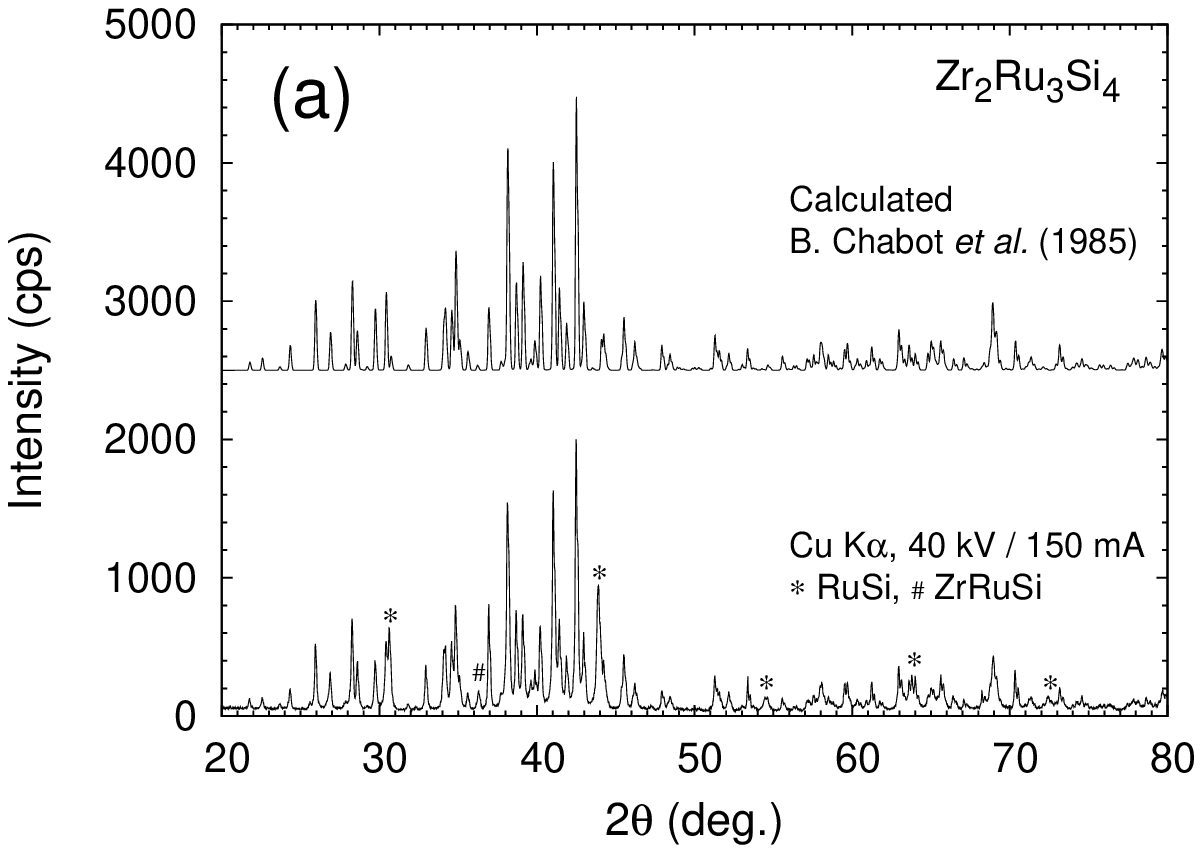}\\ 	\includegraphics[width=0.45\hsize]{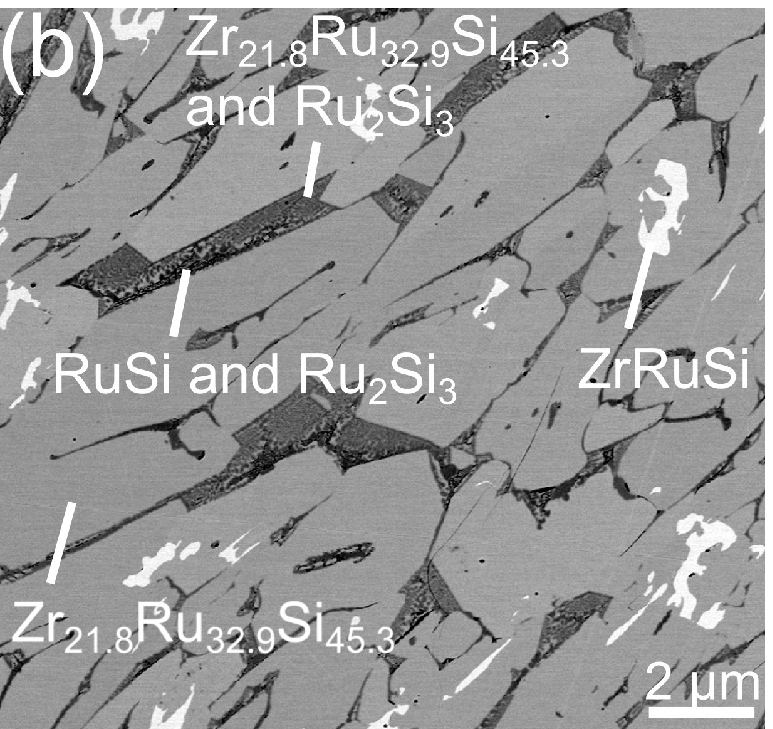}
    \caption{\label{fig1} (a) X-ray diffraction pattern of the prepared sample {\z}. Impurity peaks for RuSi and ZrRuSi are indicated by ${\ast}$ and \#, respectively. Also shown is a calculated pattern of the reported structure~\cite{Cha85}; the pattern is shifted by 2500 cps. (b) Back-scattered electron composition image of {\z}.}
 	\end{center}
\end{figure}
Figure~\ref{fig1}(a) shows the X-ray diffraction pattern of the prepared sample {\z}, compared with a calculated pattern of the structure reported by Chabot {\etal}~\cite{Cha85} Except for a few peaks, all the peaks were identified as {\z}. Lattice parameters are $a = 18.98$, $b = 5.348$, $c = 13.28$~{\AA} and $\beta = 127.73^{\circ}$, which are consistent with those reported~\cite{Cha85}. The other peaks were identified as ZrRuSi and RuSi, which indicate the existence of the contaminating phases.
Figure~\ref{fig1}(b) shows a back-scattered electron composition image of {\z}. Composition of the largest grey area
is Zr:Ru:Si = 21.8(3):32.9(4):45.3(2), which is consistent with the stoichiometric composition of {\z} within errors.
 The white phase surrounded by {\z} is ZrRuSi. The two dark areas are eutectic phases of {\z} and RuSi, and of RuSi and Ru$_2$Si$_3$. 
We note that neither RuSi nor Ru$_2$Si$_3$ is a superconductor~\cite{Busc97, Gott93}. Additionally, we verified that ZrRuSi, prepared by arc melting, does not show superconductivity down to 1.8~K as described in~\cite{Bar80, Zho86}. These impurities occupy about 10vol\% of the sample.
These results indicate that {\z} was formed as the major phase in the prepared sample.

\begin{figure}
	\begin{center}
	\includegraphics[width=0.47\hsize]{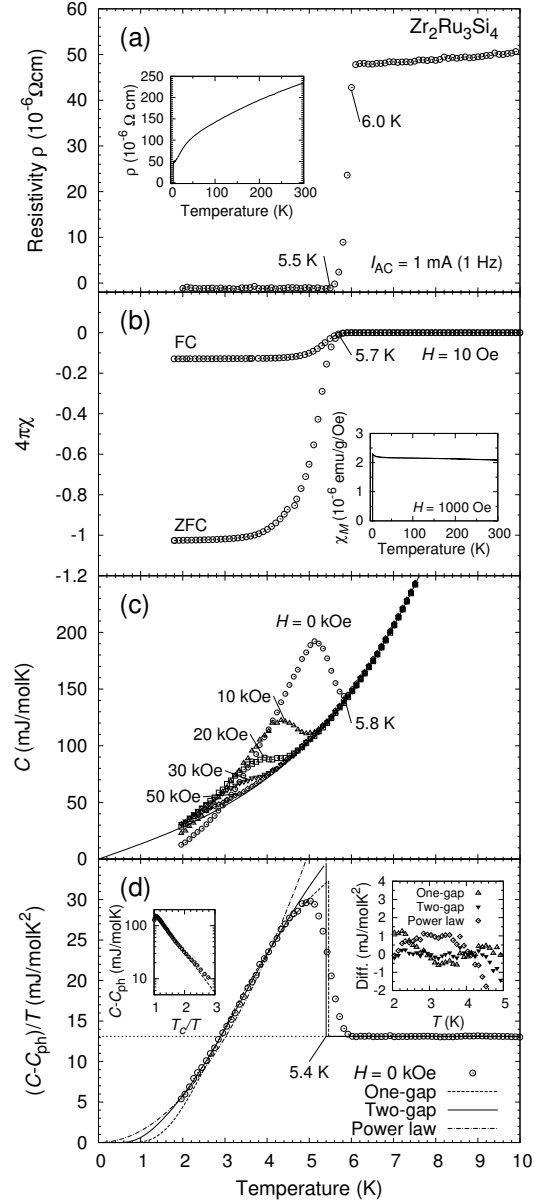}
    \caption{\label{fig2} (a) Temperature dependence of the electrical resistivity $\rho$ of {\z} at low temperatures. The inset shows $\rho$ up to $T = 300$~K. (b) Temperature dependence of the dc magnetic susceptibility $4\pi\chi$ measured with $H = 10$~Oe. FC and ZFC stand for field cooling and zero field cooling, respectively. The inset shows the temperature dependence of the mass magnetic susceptibility $\chi_M$ measured with $H = 1000$~Oe up to $T = 300$~K. (c) Temperature dependence of the specific heat $C$ under the applied magnetic field $H = 0$, 10, 20, 30 and 50~kOe. The solid line shows the best fit to $C$(0~kOe) with $C_{\rm normal} = \gamma T + \beta T^3 + \delta T^5$. (d) Phonon-subtracted specific heat $(C-C_{\rm ph})/T$ at $H = 0$~kOe. The dotted, solid and dashed-dotted lines show the one-gap, two-gap and power law fit, respectively. The left inset shows the inverse temperature dependence of $C-C_{\rm ph}$. The dotted line shows the one-gap fit. The right inset shows the differences between $(C-C_{\rm ph})/T$ and fits.} 
    \end{center}
\end{figure}
Figure~\ref{fig2}(a) shows the temperature dependence of the electrical resistivity $\rho$ of {\z} from $T = 10$ to 1.8~K. This displays the steep decreasing at 6.0~K and zero resistivity at 5.5~K, indicating the superconducting transition. 
The inset shows the high-temperature data. The resistivity at 300~K is $240 \times 10^{-6}~{\rm {\Omega}{cm}}$ and decreases with temperature going down, indicating the metallic nature of the sample. The residual resistivity ratio $\rho$(300~K)$/\rho(6.1$~K) is 4.7. 
The temperature dependence of the resistivity shows negative curvature (${\rm d}^2\rho/{\rm d}T^2 < 0$), and is very different from the Bloch-Gr\"{u}neisen theory. Such observations were made in many superconductors as in Nb$_3$Sn and may be attributed to a strong electron-phonon interaction~\cite{Wood64, Fisk76, Alle78}. Temperature dependence of the dc magnetic susceptibility $\chi$ with an applied field $H$ being 10~Oe is shown in figure~\ref{fig2}(b). $\chi$ decreases at 5.7~K and shows the large diamagnetism both in the zero field cooling (ZFC) and field cooling (FC) processes, indicating the superconducting transition of {\z}, not of impurities. The shielding volume fraction is about 100\% and flux exclusion volume is 10\%.
The inset shows mass magnetic susceptibility $\chi_M$ in high temperatures with $H = 1000$~Oe. $\chi_M$ is $2.08 \times 10^{-6}$~emu/g/Oe at 300~K and shows almost no temperature dependency, indicating the Pauli paramagnetic character. The small up-turn at low temperatures would be attributed to the magnetic impurities.
Specific heat $C$ under the applied magnetic field $H$ up to 50~kOe is shown in figure~\ref{fig2}(c). A clear jump was found below 5.8~K with $H = 0$~kOe, and it was compressed to lower temperatures by increasing $H$.
The result confirmed a bulk superconductivity with \tc = 5.5~K in {\z}.
Here, we note that other two intermetallic compounds with the Hf$_2$Ru$_3$Si$_4$-type structure, Hf$_2$Ru$_3$Si$_4$ and Yb$_2$Ru$_3$Ge$_4$, prepared by arc melting, did not show superconductivity down to 1.8~K in our magnetic susceptibility measurements (not shown).
To determine the electron and phonon contributions, the temperature dependence of normal state $C$ at $H = 0$~kOe was fitted to the function
$C_{\rm normal} = \gamma T + \beta T^3 + \delta T^5$, 
where $\gamma T$ is the electronic term, and $C_{\rm ph} = \beta T^3 + \delta T^5$ represents the phonon term. The best fitting between $6 < T < 10$~K, is shown by the solid line in figure~\ref{fig2}(c), which yields $\gamma = 13.1$~mJ/molK$^2$, $\beta = 0.279$~mJ/molK$^4$ and $\delta = 0.0011$~mJ/molK$^6$. From $\beta$, Debye temperature $\theta_D$ was estimated to be 191~K by the relation $\theta_D = \{(12\pi^4/5)(N\kb/\beta)\}^{1/3}$, where $N$ is the number of atoms, and {\kb} is the Boltzmann constant. 
The existence of the $\delta$ suggests a complex phonon density of states.
To analyze the electronic specific heat, the phonon contribution $C_{\rm ph}$ was subtracted from $C$ with $H = 0$~kOe, and temperature dependence of $(C-C_{\rm ph})/T$ was examined by the three models: one-gap $\alpha$-model, exponential function model and two-gap $\alpha$-model. $(C-C_{\rm ph})/T$ is shown in figure~\ref{fig2}(d).
Firstly, $(C-C_{\rm ph})/T$ in the superconducting state was fitted to the one-gap $\alpha$-model~\cite{Pada73} with the exponential function $C_{\rm one}(T) \propto {\exp}\left\{ \Delta(0)/{\kb}T \right\}$, where $\Delta(0)$ stands for the superconducting gap size at $T = 0$~K. The dotted line in the figure shows the fitting result. 
The left inset shows $C-C_{\rm ph}$ as a function of $\tc/T$ with semi-logarithmic scale.
The right inset shows the differences between the observed data $(C-C_{\rm ph})/T$ and fits.
The fit gives $\Delta(0)/{\kb} = 9.51$~K. $2\Delta(0)/{\kb}{\tc}$ becomes 3.5 with ${\tc} = 5.4$~K, which is compatible with the BCS value, 3.52. 
The specific heat jump ${\Delta}(C-C_{\rm ph})$ at $T = \tc$ was extrapolated for an ideally sharp transition using entropy conservation. This yields $32-13 = 19$~mJ/molK$^2$, and the normalized value becomes ${\Delta}(C-C_{\rm ph})/{\gamma}{\tc} = 1.54$, similar to that expected in the BCS weak-coupling limit of 1.43.
However, the fitting to the one-gap model shows a clear deviation below 3~K = 1/2{\tc}, indicating this is not an isotropic single-gap superconductivity.
A fit by a power law $C \propto T^3$ shown by the dash-dotted line in figure~\ref{fig2}(d), indicating point nodes in the superconducting gap, does not properly account for the experimental data, either.
Thus, we fit the data to the phenomenological two-gap $\alpha$-model~\cite{Bou01}.
In this model, the total electronic specific heat is given by the sum of the contributions of each band, $C_{\rm two}(T) \propto xC_1(T) + (1-x)C_2(T)$, where $C_1(T) = {\exp}\left\{ \Delta_1/{\kb}T \right\}$ and $C_2(T) = {\exp}\left\{ \Delta_2/{\kb}T \right\}$; they are electronic specific heat from band 1 and 2, respectively. $x$ and ($1-x$) are the relative weights for bands 1 and 2, respectively. Interband transitions due to scattering by impurities or phonons are neglected.
This model has been successfully applied to MgB$_2$~\cite{Bou01}, borocarbide superconductors~\cite{Hua06}, Lu$_2$Fe$_3$Si$_5$~\cite{Tam08}, etc.
The best fit in the two-gap model is shown by the solid line in figure~\ref{fig2}(d), yielding $\Delta_1/{\kb} = 13.7$~K, $\Delta_2/{\kb} = 5.5$~K and $x = 0.9$. The two-gap fit gives better agreement with the observed specific heat than that of the one-gap model, although this fit has some uncertainty due to the lack of the specific heat below 2~K. 
From the analysis, one can conclude {\z} not to be an isotropic single-gap superconductor, and likely to be a two-gap or anisotropic superconductor.

\begin{figure}
	\begin{center}
    \includegraphics[width=0.57\hsize]{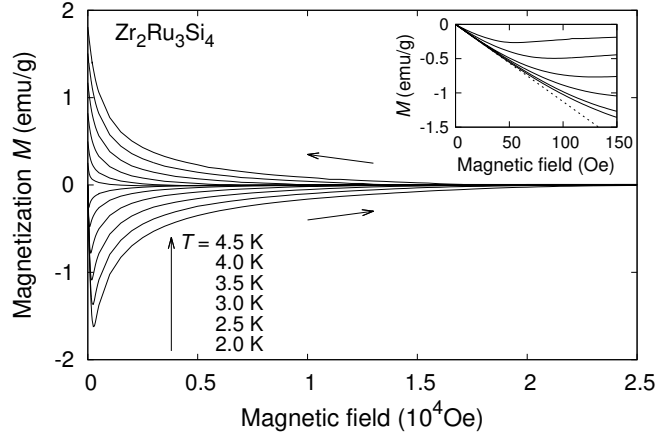}
    \caption{\label{fig3} Magnetic field dependence of the magnetization $M$ of {\z} at the constant temperatures below {\tc}. The inset shows the initial magnetization. The dotted line shows the perfect diamagnetization. } 
    \end{center}
\end{figure}
Figure~\ref{fig3} shows the magnetic field dependence of the magnetization $M$ below {\tc}. The existence of the superconducting mixed state characterises {\z} as a type-II superconductor. 
The magnetization is irreversible. This would be caused by defects and impurities in the sample; they trap magnetic flux through the sample. The inset shows the initial magnetization at low fields. The dotted line shows the perfect diamagnetization. Since the magnetization gradually deviates from the perfect diamagnetization line, the lower critical field $H_{\rm c1}(T)$ is difficult to be determined correctly. Here, we note only that $H_{\rm c1}(0)$ is approximately $1 \times 10^2$~Oe. 

\begin{figure}
	\begin{center}
    \includegraphics[width=0.57\hsize]{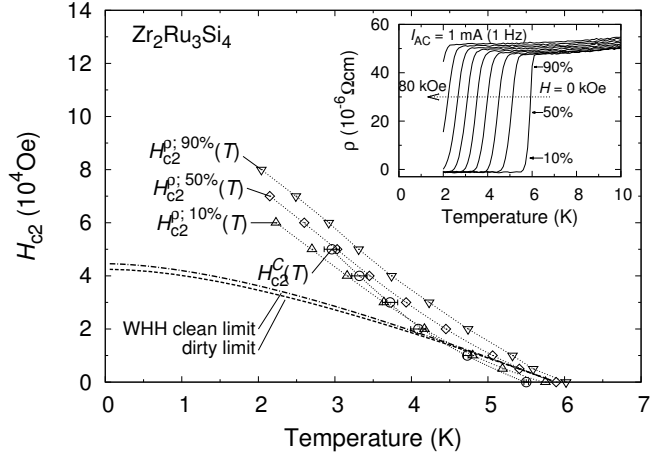}
    \caption{\label{fig4} Temperature dependence of the upper critical field {\hct}. The symbols label $H_{\rm c2}^{\rho}(T)$ for the 10\% (triangles), 50\% (squares), and 90\% (inverted-triangles) criterion of the resistivity. 
The circles stand for $H_{\rm c2}^{C}(T)$ taken from the heat capacity.
The dotted lines shows guide to the eyes. The dashed-dotted and broken lines show WHH curves with clean and dirty limit, respectively.
The inset shows the temperature dependence of the electrical resistivity under the magnetic field from 0 to 80~kOe. }
    \end{center}
\end{figure}
Temperature dependence of the upper critical field {\hct} is shown in figure~\ref{fig4}. {\hct} is estimated based on the electrical resistivity $H_{\rm c2}^{\rho}(T)$ and heat capacity $H_{\rm c2}^{C}(T)$ under the magnetic field. 
The electrical resistivity under the magnetic field up to 80 kOe is illustrated in the inset. The transition temperature is suppressed by the field. $H_{\rm c2}^{\rho}(T)$ was determined by 10, 50 and 90\% of the normal state resistivity before the transition. 
$H_{\rm c2}^{C}(T)$ was taken from the centre of the specific heat jump in figure~\ref{fig2}(c).
The curvature between $H_{\rm c2}^{\rho}(T)$ and $H_{\rm c2}^{C}(T)$ displays a small difference. This would be because resistivity measurements are easily affected by grain boundaries compared to capacity measurements.
{\hct} shows the sizable positive curvature near {\tc}. 
The gradient of the curve in the vicinity of {\tc}, $-{\rm d}H_{\rm c2}^{\rho; 50\%}/{\rm d}T |_{T \sim {\tc}}$, is 10 kOe/K. 
The Werthamer-Helfand-Hohenberg (WHH) curves with clean and dirty limit~\cite{Hel66}, which based on the isotropic single-gap model, are plotted by the dashed-dotted and broken lines, respectively. In conventional superconductors, {\hct} is well described by WHH curves.  
It is clear that the observed data are inconsistent both with the WHH curves, indicating that the superconducting gap symmetry of {\z} is not an isotropic $s$-wave. If $H_{\rm c2}^{\rho; 50\%}$ is extrapolated linearly to 0~K, $H_{\rm c2}(0)$ is about $1 \times 10^2$~kOe. The coherence length $\xi$ estimated by the relation $\xi = \left\{ \phi_0 / 2 \pi H_{\rm c2}(0) \right\}^{1/2}$, is 6~nm, where $\phi_0$ is the flux quantum.

\begin{figure}
	\begin{center}
    \includegraphics[width=0.47\hsize]{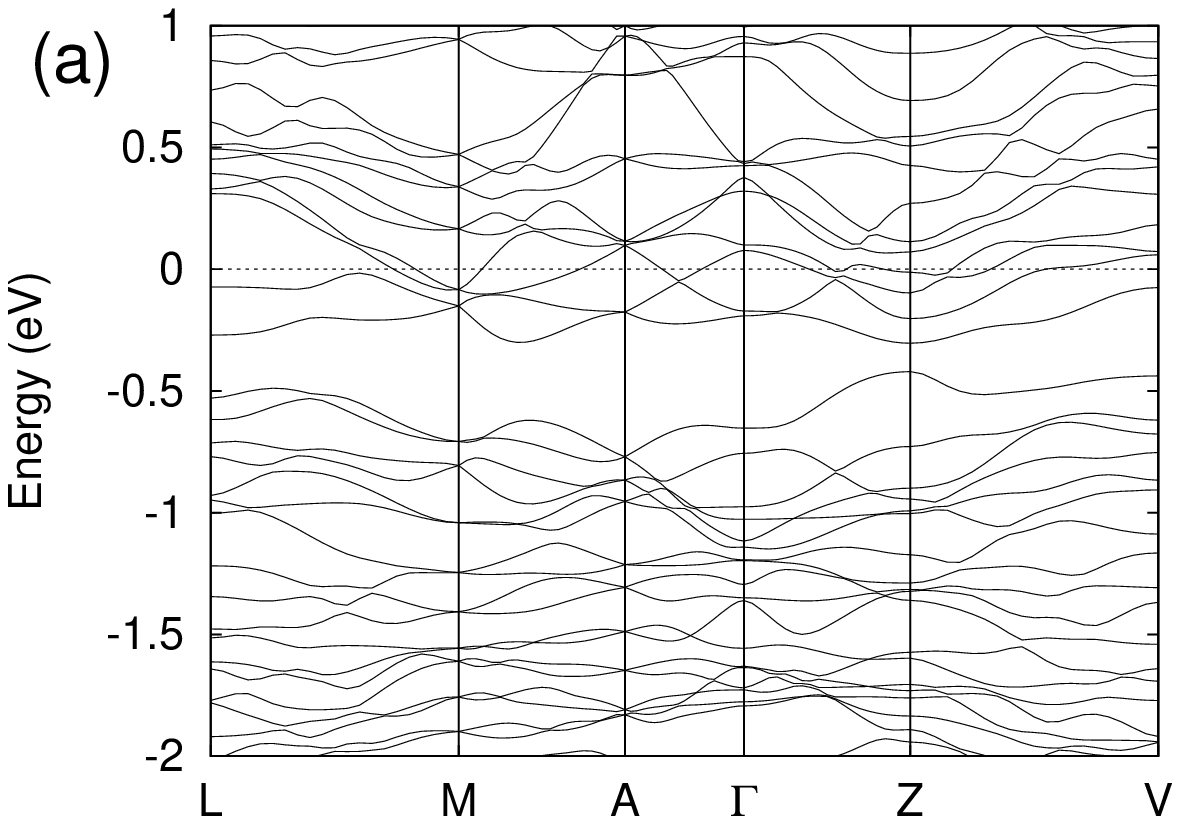}\\     \includegraphics[width=0.47\hsize]{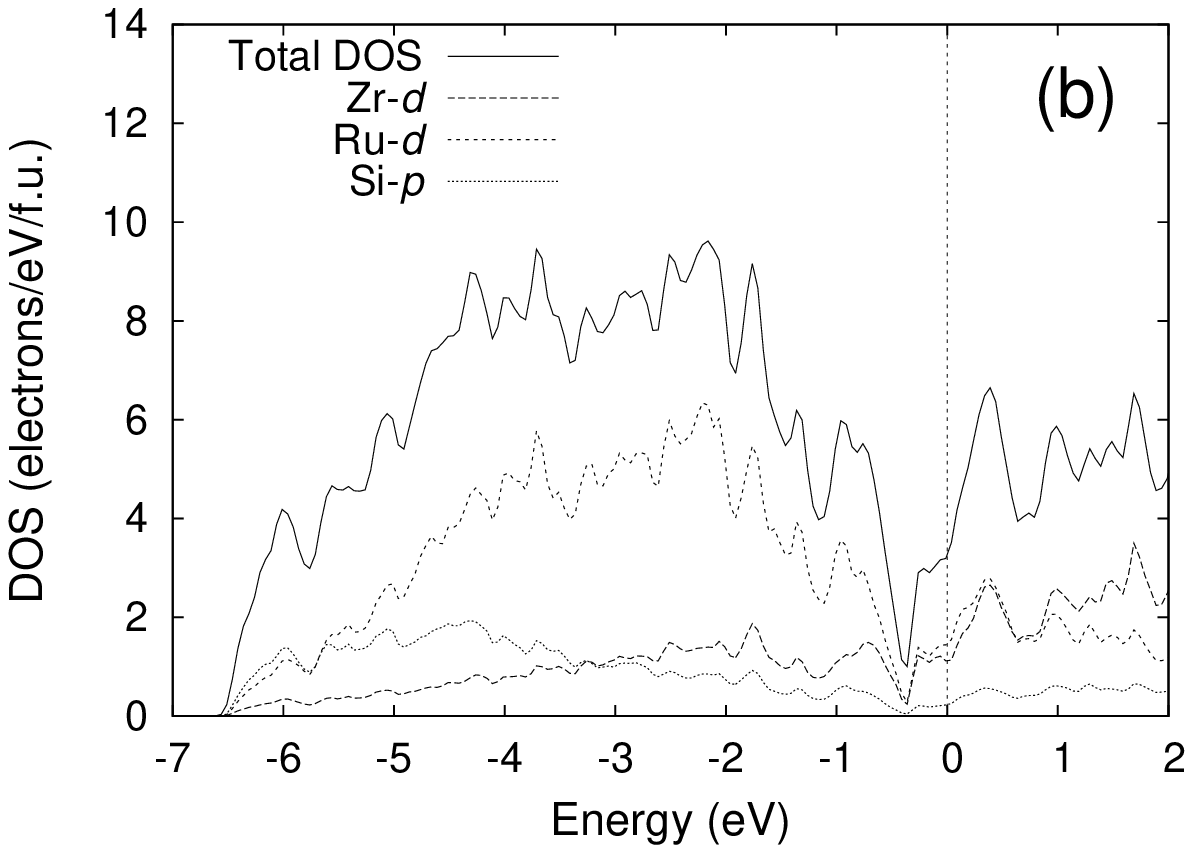}\\
    \includegraphics[width=0.47\hsize]{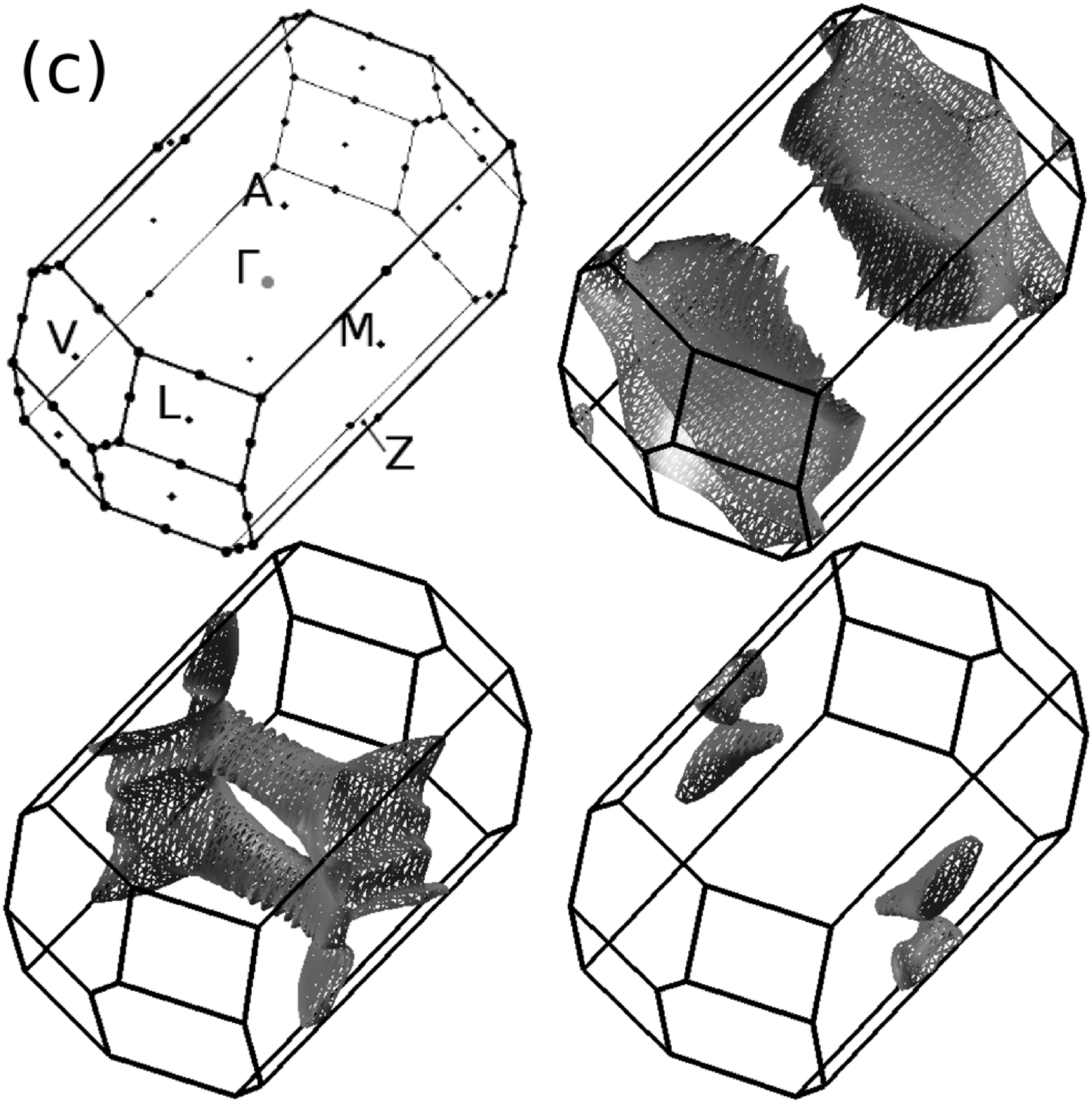}\\
    \caption{\label{fig5} (a) Band structure for {\z} in an energy window near Fermi energy ($E_F$). $E_F$ is set to 0~eV. (b) Density of states of {\z} with $E_F = 0$~eV. (c) Fermi surfaces (FSs) shown in the first Brillouin zone centred at $\Gamma$ point~\cite{Kok03}. High symmetry points (upper left). Hole FS around V point (upper right). Electron FSs around M and V points (lower left and right).}     \end{center}
\end{figure}
Figure~\ref{fig5}(a) shows band structure for {\z} obtained by first-principle's calculations in an energy window near Fermi energy $E_F$. Figure~\ref{fig5}(b) illustrates DOS of {\z} in the unit of electrons/eV/formula-unit (f.u.). In both the figures, $E_F$ is set to 0 eV. 
Density of states at $E_F$ is $N(E_F) = 3.2$~electrons/eV/f.u.
Partial DOS of each element is also shown in figure~\ref{fig5}(b). There is little difference between sites in each atom (not shown), indicating a highly itinerant electron character of {\z}. The partial DOS indicates that the bands near $E_F$ consist of Ru $4d$, Zr $4d$ and Si $3p$ states. The electrons of all the three species of ions contribute to form the Fermi surfaces (FSs) and likely the superconductivity as well.
In the lower bands from -6.5 to -0.5~eV, Ru $4d$ state is relatively predominant. 
FSs plotted in the first Brillouin zone are illustrated in figure~\ref{fig5}(b)~\cite{Kok03}. The two electron FSs exist around M and Z points, and a hole FS around V point. The anisotropic FSs reflect the low-symmetric crystal structure of {\z}.

\begin{figure}
	\begin{center}
	\includegraphics[width=0.57\hsize]{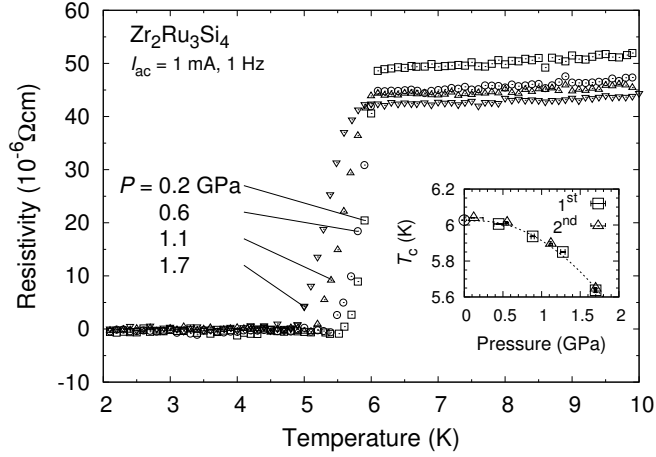}
    \caption{\label{fig6} Temperature dependence of the electrical resistivity under the pressure $P$ of 0.2, 0.6, 1.1 and 1.7 GPa. Pressure dependence of {\tc} is shown in the inset. The squares and triangles stand for the $1^{\rm st}$ and $2^{\rm nd}$ set of measurements, respectively. {\tc} at 0~GPa is obtained from a resistivity measurement without a high-pressure cell. The dotted line shows the fit with a quadratic function.} 
    \end{center}
\end{figure}
Temperature dependence of the electrical resistivity under the pressure $P = 0.2$, 0.6, 1.1 and 1.7~GPa is shown in the figure~\ref{fig6}. {\tc} is decreased with increasing the pressure, as seen in the majority of the superconductors.
The inset shows the pressure dependence of {\tc}. {\tc} is determined by the intersection of the extrapolated resistivity lines above and below the upper side of the transition.
The pressure dependence of {\tc} was fitted with a quadratic function and gives ${\tc} = -0.16P^2 + 0.06P + 6.02$.
According to the BCS theory, a relation between {\tc}, typical phonon energy $\langle\omega\rangle$ and the interaction strength $N(0)V$ is given by 
${\tc} \sim 1.14 \langle\omega\rangle \exp \left\{-1/N(E_F)V\right\}$, 
where $V$ is the pairing potential arising from the electron-phonon interaction. 
By the McMillan's expression~\cite{McM68}, $N(E_F)V = N(E_F) \langle I^2 \rangle / M\langle \omega^2 \rangle$, the relation becomes,
\begin{equation}
{\tc} \sim 1.14 \langle\omega\rangle \exp \left\{ \frac{-M\langle \omega^2 \rangle}{N(E_F)\langle I^2 \rangle} \right\}, 
\end{equation}
where $\langle I^2 \rangle$ stands for the average square electronic matrix element, $\langle \omega^2 \rangle$ the average square phonon frequency, $M$ the mass of the ion.
{\tc} is more affected by $\omega$ in the exponent term than in the coefficient term; an increase of $\omega$ results in a decrease of {\tc}.
Therefore, by the representation, 
the pressure dependence result indicates that the pressure-induced stiffening of the lattice vibration spectrum dominates over the minor changes in the electronic properties in {\z}~\cite{Hop71, Schi07}. 
Additionally, the result suggests that the contraction of the volume is disadvantageous to the superconductivity of {\z}. This may explain that Hf$_2$Ru$_3$Si$_4$ is not a superconductor; the cell volume of Hf$_2$Ru$_3$Si$_4$ is 1\%
 smaller than that of {\z}~\cite{Cha85}.

\section{Discussion}
The zero resistivity, magnetic flux expulsion and heat-capacity jump clearly demonstrate superconductivity of {\z} below $T = 5.5$~K. In addition, the magnetic field dependence of the magnetization suggests that it is type-II superconductivity. This can be verified numerically with the Ginzburg-Landau parameter $\kappa$.
$\kappa$ is estimated by the relation $H_{\rm c2}(0)/H_{\rm c1}(0) = 2{\kappa}^2/{\rm ln}{\kappa}$. Employing the approximate values of $H_{\rm c2}(0) = 1 \times 10^2$~kOe and $H_{\rm c1}(0) = 1 \times 10^2$~Oe gives $\kappa = 5 \times 10^1$. The fact that $\kappa$ is quite larger than $1/\sqrt{2}$ also supports the type-II superconductivity.

An empirical electron-phonon interaction strength $\lambda_M$ is estimated by the McMillan formula~\cite{McM68},
\begin{equation}
\lambda_M = \frac{1.04 + \mu^{\ast}\ln(\theta_D/1.45 \tc)}{(1-0.62\mu^{\ast}\ln(\theta_D / 1.45\tc) -1.04},
\end{equation}
where the repulsive screened Coulomb part $\mu^{\ast}$ is set to 0.13. This yields $\lambda_M = 0.77$, characterising {\z} as an intermediate-coupling superconductor.
On the other hand, $N(E_F)$ and $\gamma$ give $\lambda_N = 0.74$ by the relation $(1+\lambda_N) = 3\gamma/(\pi^2{\kb}^2)/N(E_F)$; this is consistent with $\lambda_M$.

Additionally, two results imply that the superconducting gap is not an isotropic single-gap. One is the deviation of the electronic heat capacity from the BCS $s$-wave model below 1/2{\tc}. The other is the positive curvature of {\hct} near {\tc}.
The similar behaviours were observed in MgB$_2$~\cite{Bou01}, YNi$_2$B$_2$C~\cite{Shu98}, Lu$_2$Fe$_3$Si$_5$~\cite{Tam08}, Sc$_5$Ir$_4$Si$_{10}$~\cite{Tam08}, $\beta$-Bi$_2$Pd~\cite{Ima12}, La$_3$Pd$_4$Si$_4$~\cite{Kne13b}, etc.; they are considered to be multi-gap superconductors. 
Except for the multi-gap scenario, the positive curvature of {\hct} is suggested theoretically according to an anisotropic gap~\cite{Met97}, quasi-2D fluctuations~\cite{Bah95}, quantum critical point scenario~\cite{Kot96} and bipolaronic scenario~\cite{Ale93}. 
In {\z}, multi-gap or anisotropic gap scenario would be preferable, since {\z} has anisotropic and more than one Fermi surfaces.
For further understanding of the superconducting gap symmetry of {\z}, {\hct} and heat capacity measurements on a single crystal and in lower temperatures below $T = 2$~K are required. 

\section{Conclusions}
In this study, superconducting properties of the polycrystalline {\z} were investigated by measuring electrical resistivity, magnetization and heat capacity. Superconducting transition of ${\tc} = 5.5$~K was confirmed. {\z} was found to be a type-II and intermediate-coupling superconductor. 
Interestingly, the deviation of the heat-capacity from the BCS isotropic $s$-wave curve and the positive curvature of {\hct} near {\tc} was found; the similar phenomena are often seen in multi-gap or anisotropic-gap  superconductors. 
Additionally, the first principle's calculation shows that {\z} has plural anisotropic Fermi surfaces.
These facts indicate that {\z} is not an isotropic single-gap superconductor, but probably a multi-gap or anisotropic gap superconductor.
{\z} supplies a new example of non-isotropic-single-gap superconductors with monoclinic structural symmetry.
 
\ack
We thank H. Murata for valuable comments on the first-principle's calculation, and H. Mamiya, H. Kitaguchi and H. Kitazawa for lending us their PPMS and MPMS. This work was supported by the Funding Program for World-Leading Innovative R\&D on Science and Technology (FIRST), Japan.

\section*{References}
\bibliography{ZrRuSi}

\end{document}